\documentclass[aps,prd,superscriptaddress,amsfonts,amssymb,amsmath,showpacs,10pt,onehalfspacing]{revtex4-2}
\usepackage[utf8]{inputenc}
\usepackage{siunitx}
\usepackage{url}
\usepackage{tensor}
\usepackage{longtable}
\usepackage{array}
\usepackage{mathtools}
\usepackage{enumitem}
\usepackage{amssymb}
\usepackage{amsmath}
\usepackage{makecell,tabularx,multirow}
\usepackage{ulem}
\usepackage{color, colortbl}
\usepackage{adjustbox}
\usepackage{tikz}
\usetikzlibrary{positioning}
\usetikzlibrary{shapes.geometric, arrows}
\usepackage[pages=some]{background}
\usepackage{hyperref}
\usepackage{nameref}
\usepackage{accents}
\usepackage{todonotes}
\usepackage{eqnarray}
\usepackage{makecell}

\usepackage{makecell}
\hypersetup{
    colorlinks=true,
    linkcolor=blue,
    filecolor=blue,
    urlcolor=blue,
    citecolor=blue
}
\usepackage{comment}
\usepackage{booktabs}

\newcolumntype{C}{>{\centering\arraybackslash}X}
\newcolumntype{x}[1]{>{\centering\arraybackslash\hspace{0pt}}p{#1}}

\tikzstyle{startstop} = [rectangle, rounded corners, 
text width=7em, minimum height=1cm,text centered, draw=black, fill=red!30]
\tikzstyle{io} = [trapezium, trapezium left angle=70, trapezium right angle=110, 
text width=7em, minimum height=1cm, text centered, draw=black, fill=blue!30]
\tikzstyle{process} = [rectangle, 
text width=7em, minimum height=1cm, text centered, draw=black, fill=orange!30]
\tikzstyle{decision} = [diamond, minimum height=1cm, text centered, text width=5.5em, node distance=3cm, draw=black, fill=green!30]
\tikzstyle{arrow} = [thick,->,>=stealth]

\begin{document}

\title{Magnetar driven Dainotti relation with up to 56\% reduced uncertainties: probing the Expansion History of the Universe}

\author{Konstantinos F. Dialektopoulos}
\email{kdialekt@gmail.com}
\affiliation{Institute of Applied Sciences, The Malta College of Arts, Science and Technology, Paola, Malta}
\affiliation{Institute of Space Sciences and Astronomy, University of Malta, Msida, Malta}
\affiliation{Department of Physics, University of Malta, Msida, Malta}

\author{Aleksander Łukasz Lenart}
\email{aleksander.lenart@doctoral.uj.edu.pl}
\affiliation{Jagiellonian University, Doctoral School of Exact and Natural Sciences, Kraków, Poland}
\affiliation{Astronomical Observatory of Jagiellonian University in Kraków, Orla 171, 30-244 Kraków}

\author{Maria Giovanna Dainotti}
\email{maria.dainotti@nao.ac.jp }
\affiliation{Division of Science, National Astronomical Observatory of Japan, 2-21-1 Osawa, Mitaka, 181-8588 Tokyo, Japan}
\affiliation{The Graduate University for Advanced Studies (SOKENDAI), Shonankokusaimura, Hayama, Miura District, Kanagawa 240-0115, Japan}
\affiliation{Space Science Institute, 4765 Walnut St Ste B, Boulder, CO 80301, USA}
\affiliation{Nevada Center for Astrophysics, University of Nevada, 4505 Maryland Parkway, Las Vegas, NV 89154, USA}

\author{Jackson Levi Said}
\email{jackson.said@um.edu.mt}
\affiliation{Institute of Space Sciences and Astronomy, University of Malta, Msida, Malta}
\affiliation{Department of Physics, University of Malta, Msida, Malta}

\author{Nissim Fraija}
\affiliation{Instituto de Astronomía, Universidad Nacional Autónoma de México, Circuito Exterior, C.U., A.P. 70-264, Mexico}

\begin{abstract}
The persistent tension between early- and late-Universe estimations of the Hubble constant, together with recent DESI hints of a dynamical dark-energy equation of state, motivates the search for cosmological probes that extend the distance ladder beyond the reach of Type Ia supernovae (SNe~Ia) and Cosmic Chronometers (CC), into the largely unexplored redshift range $z \sim 2$--$6$. 
We use the X-ray plateau emission of long gamma-ray bursts (GRBs), standardized through the two-dimensional Dainotti relation between plateau luminosity and rest-frame plateau duration, as a probe of this gap. We analyze two GRB samples with different selection philosophies: the platinum sample (50 GRBs, selected via light-curve quality cuts), hereafter the P50 sample; and the magnetar sample (23 GRBs, selected by requiring the magnetar spin-down model to be statistically preferred over the empirical Willingale parametrization), hereafter the M23 sample. 
Initially, we use a two-step simultaneous-fitting procedure, anchored independently to the Master sample SNe~Ia compilation and to Cosmic Chronometers, we calibrate the Dainotti relation and propagate the result into a joint flat-$\Lambda$CDM cosmological inference, sampled via MCMC with \texttt{cobaya}. 
Although the magnetar sample contains less than half the GRBs of the platinum sample, its intrinsic scatter is $\sim 41\%$ smaller, indicating that the physically justified, robust selection is a better strategy than the quality criteria in the sampling optimization. 
Despite this difference, the two samples yield cosmological constraints that are almost identical in $H_0$ and $\Omega_m$ once combined with SNe~Ia and CC. Specifically, M23 gives $H_0 = 69.45\pm0.32\ \mathrm{km\,s^{-1}\,Mpc^{-1}}$ and $\Omega_m = 0.34 ^{+0.01}_{-0.02}$ while P50 gives $H_0 = 69.46\pm0.33\ \mathrm{km\,s^{-1}\,Mpc^{-1}}$ and $\Omega_m = 0.34\pm0.02$. 
Thus, the magnetar sample values are in $3$--$3.5\sigma$ tension with both Planck and SH0ES. 
Therefore, we conclude that the scatter of the relation plays a more important role in the cosmological constraints than the sample size. Furthermore, we developed a novel method of simultaneous fitting of the Dainotti relation and light curves, which allows for a $\sim 56\%$ reduction in intrinsic scatter. 
Although GRBs still cannot provide standalone constraints on the cosmological parameters, we demonstrate a method that can improve the constraining efficiency of the future, larger samples. These results establish magnetar-selected GRB plateaus as a robust, high-redshift complement to existing distance-ladder probes.
\end{abstract}

\maketitle

\section{Introduction}

The standard cosmological model, flat $\Lambda$CDM, has provided a remarkably successful framework for describing the large-scale structure and evolution of the Universe over several decades of precision measurements. Yet a growing body of observational evidence is placing its foundational assumptions under increasing stress \cite{CosmoVerseNetwork:2025alb}. The most persistent of these challenges is the Hubble tension: the statistically significant discrepancy between the value of $H_0$ inferred from the early Universe via the Cosmic Microwave Background~\cite{Planck:2018vyg} and the direct local measurements anchored to the Cepheid distance ladder~\cite{Riess:2021jrx}, which currently stands at approximately $5\sigma$. The Hubble tension is becoming more and more at the center of debate as additional trends of $H_0$ as a function of redshift have been revealed \citep{Dainotti2021hubble,Dainotti2022hubble,DESIMONE2024}. Equally notable are the recent results from the Dark Energy Spectroscopic Instrument (DESI), whose baryon acoustic oscillation measurements across multiple redshift bins show a preference for a dynamical dark energy equation of state over a cosmological constant at the $2-3\sigma$ level~\cite{DESI:2024mwx}, independently motivating a reexamination of the $\Lambda$CDM paradigm.

A common thread running through these tensions is the limited redshift coverage of the probes that anchor our cosmological measurements. Type Ia Supernovae (SNe~Ia), the workhorse of the late-time distance ladder, reach up to $z \lesssim 2.9$ in the Master SNe compilation~\cite{Dainotti2025JHEAp..4800405D} (however, very few sources have $z\geq 1.5$), while Cosmic Chronometers (CC) provide direct $H(z)$ measurements up to $z \approx 2$~\cite{Moresco:2016mzx, Moresco:2020fbm}. The redshift range between these probes and the CMB last scattering surface at $z \approx 1100$ remains largely unexplored by independent distance indicators, yet it is precisely this regime that is most sensitive to the dynamics of dark energy and the growth of structure. Extending the cosmic distance ladder into this unexplored territory is therefore a priority for modern observational cosmology.

Gamma-Ray Bursts (GRBs) are among the most energetic transient phenomena in the Universe, detectable up to $z \approx 9.4$~\cite{Cucchiara:2011pj} and potentially beyond with future facilities~\cite{Lamb:1999qv}. This extreme redshift reach makes them uniquely positioned to probe the expansion history of the Universe in the redshift gap between SNe~Ia and the CMB. The key challenge in using GRBs as standard candles is the absence of a physical calibration analogous to the Cepheid period-luminosity relation: GRB energetics span several orders of magnitude, and deriving a usable distance indicator requires identifying empirical correlations among observable properties that are sufficiently tight and cosmology-independent to be useful. Several such correlations have been proposed~\cite{Amati:2002ny,Yonetoku:2003gi,Schaefer:2006pa}, but the most promising for cosmological applications are those involving the X-ray plateau phase of GRB afterglows, observed by the Neil Gehrels \textit{Swift} Observatory.

The plateau phase arises from sustained energy injection into the GRB afterglow, most naturally interpreted as spin-down radiation from a newly formed millisecond magnetar~\cite{DallOsso:2010uxj,Rowlinson:2014dja,Rea2015,Stratta2018}. The two-dimensional Dainotti relation~\cite{Dainotti:2008vw, Dainotti:2010ki} connects the rest-frame plateau end time $T^*_a$ to the plateau luminosity $L_a$, and has been shown to be independent of the assumed cosmology and selection biases and redshift evolution tested through the Efron-Petrosian method~\cite{Dainotti:2013fra}. In this work, we make use of two GRB samples that differ in their selection strategy. The platinum sample~\cite{Dainotti:2020azn} applies the most stringent quality cuts: no flares, flat plateau profiles, and temporal indices consistent with the energy injection model both during and after the plateau, yielding $50$ GRBs spanning $z \in [0.553, 5.000]$. Hereafter, we call this sample P50. The magnetar sample, defined here, applies a broader magnetar-motivated selection, resulting in a set of $23$ GRBs spanning $z \in [0.624, 5.570]$. Hereafter, we call this sample M23.

Despite comprising less than half the number of GRBs, M23 exhibits an intrinsic scatter in the Dainotti relation that is already $\sim 41\%$ smaller than that of P50 when both are treated with the standard two-step calibration approach, indicating that the magnetar-engine selection removes a source of physical dispersion that generic quality cuts alone cannot eliminate. When M23 is instead calibrated through the novel joint method introduced in this work — simultaneously fitting the light-curve parameters, the Dainotti correlation, and the cosmological parameters — the reduction in intrinsic scatter relative to P50 increases to $\sim 56\%$. The comparison between the two samples therefore provides an empirical test of whether selection criteria, sample size, or calibration methodology is the dominant factor in determining the cosmological constraining power of GRBs.

A fundamental obstacle in using GRBs as cosmological probes is the circularity problem: the conversion of observed fluxes and timescales into luminosities requires a distance, which in turn depends on the cosmological model one wishes to constrain. Several strategies have been developed to break this circularity, ranging from simultaneous fitting of correlation and cosmological parameters~\cite{Dainotti:2013fra} to model-independent calibration via Gaussian processes~\cite{Dainotti:2022wli} or artificial neural networks~\cite{Dialektopoulos:2023dhb}. In this work, we adopt the two-step simultaneous fitting approach: in the first step, the Dainotti parameters are determined at low redshift jointly with the cosmological parameters using an external low-z sample; in the second step, the calibrated parameters are used as Gaussian priors in a GRB-only cosmological inference. We apply two independent calibrators, the SNe~Ia master sample and the CC compilation, and perform every analysis independently for both GRB samples, demonstrating that the resulting cosmological constraints are robust to both the choice of calibrator and the choice of GRB selection criteria.

Beyond the baseline analysis, we investigate a systematic extension motivated by the structure of the GRB likelihood. We introduce a redshift evolution correction to the Dainotti relation, allowing the intercept to vary with $(1+z)$ and testing whether unaccounted-for evolution biases the cosmological parameter estimates. What we find is that the baseline results are stable, strengthening confidence in the robustness of the cosmological inference.

This paper is organized as follows. Section~\ref{sec:method} describes the cosmological framework, the datasets, and the likelihood construction. Section~\ref{sec:results} presents the results of the calibration runs, the cosmological parameter estimation under all probe combinations, and the systematic robustness tests. We summarize our findings and discuss their implications in Section~\ref{sec:conclusions}.

\section{Methodology}
\label{sec:method}

In this section we lay out the methodology behind our GRB-based cosmological probe and its combination with independent low-redshift datasets. We first describe the physical model underlying magnetar-powered plateau emission and the criteria used to isolate a clean sample of well-behaved GRB light curves. We then introduce a joint fitting procedure that simultaneously constrains individual light-curve parameters and the Dainotti correlation itself, which mitigates the impact of sparse or gapped data on the fit. Finally, we specify the background cosmology and describe the datasets, the Master SNe~Ia compilation, cosmic chronometers, and two independent GRB samples, that enter the combined likelihood analysis.

\subsection{Data}
\label{sec:data}

\paragraph{Type Ia Supernovae.}
We use the master sample~\cite{Dainotti2025JHEAp..4800405D}, comprising 3714 SNe~Ia spanning the redshift range $z \in [0.00122, 2.261]$. Each data point provides an observed distance modulus $\mu_{\rm obs}$ and its uncertainty. The full $3714 \times 3714$ covariance matrix $\mathbf{C}_{\rm SN}$, accounting for both statistical and systematic uncertainties including calibration, mass-step corrections, and peculiar velocity contributions, is used in the likelihood. This is the same compilation used in~\cite{Dialektopoulos:2023dhb}, ensuring direct comparability with our previous results.

\paragraph{Cosmic Chronometers.}
Cosmic Chronometers (CC) are passively evolving early-type galaxies whose differential age evolution provides a direct, cosmology-independent measurement of the Hubble parameter at discrete redshifts via $H(z) = -(1+z)^{-1}\,\mathrm{d}z/\mathrm{d}t$~\cite{Jimenez:2001gg}. We use the compilation of 32 measurements spanning $z \in [0.07, 1.965]$~\cite{Moresco:2016mzx, Moresco:2020fbm}, each providing $H_{\rm obs}(z_i)$ and its uncertainty $\sigma_{H,i}$. The diagonal covariance structure is adopted following standard practice in the literature~\cite{Yu:2017iju, Gomez-Valent:2018hwc}, noting that the full Moresco covariance matrix~\cite{Moresco:2020fbm} introduces correlated systematics from stellar population modeling that we do not include here but which represent a natural extension of this analysis.

\paragraph{Gamma-Ray Bursts.}
We make use of two GRB samples which obey the Dainotti relation, drawn from observations by the Neil Gehrels \textit{Swift} Observatory, both based on the plateau emission phase of X-ray afterglows. Using two independent samples with different selection criteria allows us to assess the robustness of the cosmological results against the choice of GRB dataset, and the comparison between them constitutes an additional consistency test of the analysis. We analyzed all data observed until January 2026; the youngest GRB in our sample that fulfilled all selection criteria was observed in July 2025.

The first sample is the Magnetar compilation, comprising $23$ long GRBs spanning the redshift range $z \in [0.624,\, 5.570]$, selected on the basis of well-characterized X-ray plateau emission consistent with the magnetar spin-down energy injection model (see Sec. \ref{sec:magnetar_sample}). The second is the platinum sample~\cite{Dainotti:2020azn}, a filtered set of $50$ GRBs covering $z \in [0.553,\, 5.000]$, selected by imposing stricter quality criteria: no flares during the plateau phase, at least 5 data points in its beginning, and a relatively flat plateau. This selection significantly reduces the intrinsic scatter of the Dainotti relation at the cost of a more conservative data selection, making the two samples complementary rather than redundant. The platinum sample is selected primarily through light-curve quality and morphology criteria, whereas M23 uses a magnetar-motivated model-selection criterion.

For each GRB in both samples, the dataset provides the observed plateau flux $F_a$ and its uncertainty $\sigma_{\log F_a}$, the observed plateau end time $T_a$ and its uncertainty $\sigma_{\log T_a}$, the spectral $K$-correction \citep{Bloom2001} $K_a$ accounting for the transformation between the observer-frame and rest-frame energy bands, and the spectroscopic redshift $z$. The rest-frame plateau end time is obtained as $\log T^*_a = \log T_a - \log(1+z)$. Both samples extend the distance ladder to redshifts well beyond the reach of both SNe~Ia and CC data, with the magnetar and platinum samples reaching $z \approx 5.6$ and $z \approx 5.0$ respectively, making GRBs the primary high-redshift component of the combined analysis.

\subsection{Magnetars as cosmological probes}
\label{sec:magnetar_sample}

Distance-ladder-type cosmological probes are characterized by the fact that distance can be measured without knowledge of redshift. For example, in the case of Cepheid stars, this is achievable due to the existence of a correlation between absolute luminosity and the period of their variability. For GRBs, researchers rely on correlations such as the \cite{Amati:2002ny} correlation between the isotropic equivalent energy of the prompt emission and the photon energy at the peak of the spectrum, or the \cite{Dainotti:2008vw} correlation between the plateau luminosity ($\log_{10}L_a$) and the source rest-frame plateau duration ($\log_{10}T_a^*$). However, GRBs are much harder to observe with high quality than probes such as Cepheid stars or Type Ia supernovae. Their transient nature makes them difficult to detect. Additionally, GRB samples are, in general, composed of sources of several different origins. Their progenitor can be either the core collapse of a very massive star (collapsar) \citep{mazets1981catalog,kouveliotou1993identification,paczynski1998ApJ...494L..45P, Woosley2006ARA} or the merger of two neutron stars \citep{1992ApJ...395L..83N, 1992ApJ...392L...9D, 1992Natur.357..472U, 1994MNRAS.270..480T, 2007PhR...442..166N, 2017ApJ...846L...5G, Goldstein2017}. Both types of events can result in the creation of either a black hole or a magnetar, and both compact objects can act as a central engine for a GRB, powering its light curve over an extended period and producing the feature known as the plateau. Additionally, both models predict a correlation between the luminosity at the end of the plateau and its duration \cite{Rowlinson:2014dja,Stratta:2018xza, Cannizzo2009,Cannizzo2011, Lenart2025}.

\begin{figure}
    \centering
    \includegraphics[width=0.95\linewidth]{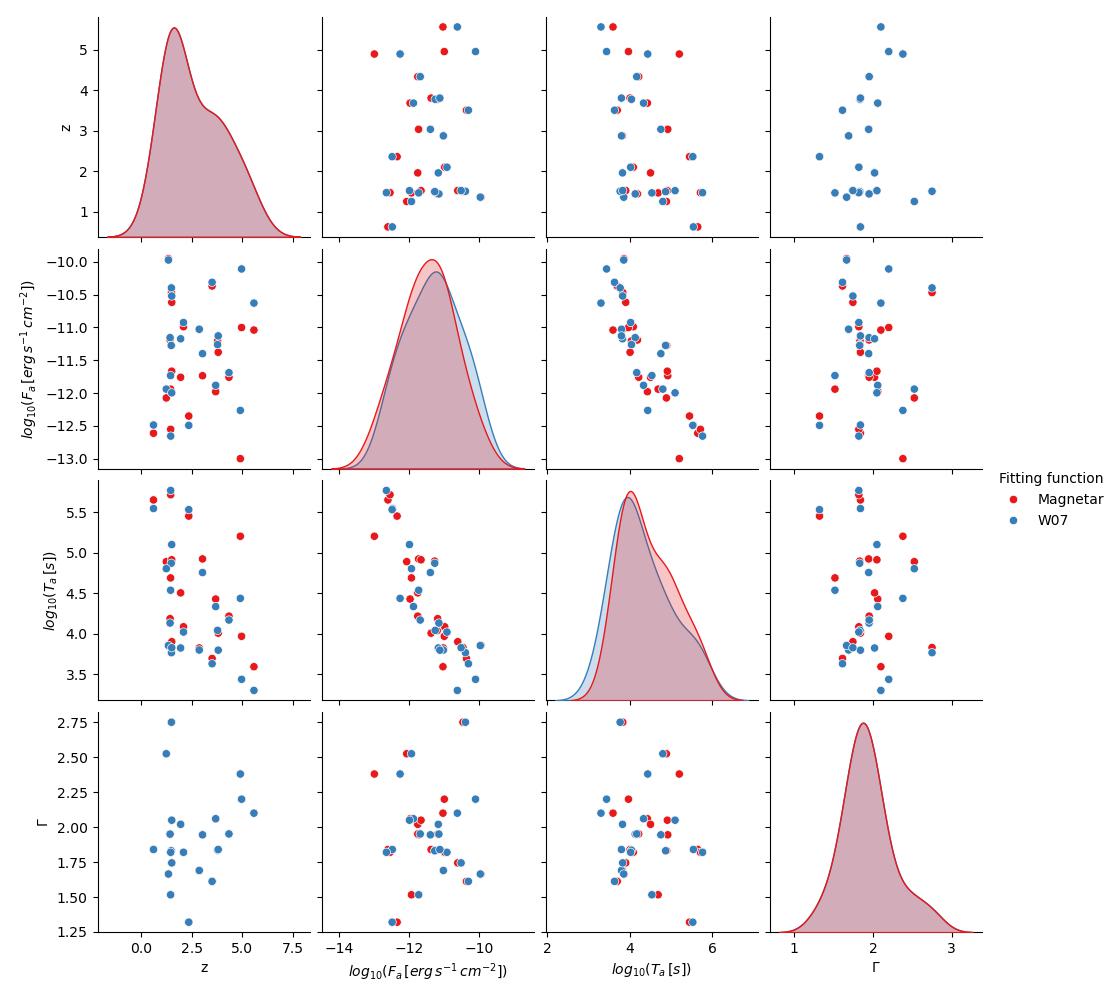}
    \caption{The parameter space of a selected sample of 23 magnetar-origin GRBs. The flux at the end of the plateau and its duration were obtained by fitting ideal magnetar and W07 models, which are shown as red and blue dots, respectively. The photon index $\Gamma$ and redshift $z$ values are the same irrespective of the fitted model.}
    \label{fig:Magnetar_matrix}
\end{figure}

Within the magnetar model, a newly born, rapidly rotating neutron star loses rotational energy through electromagnetic torques associated with its strong magnetic field, providing a sustained power source for the plateau emission \cite{DaiLu1998A&A,DaiLu1998PhRvL,zhangmeszaros2001,Zhang2006,LuZhang2014,Lu2015}.  In the case of a black hole created in the collapsar, the light curve either mimics the accretion rate of the fallback material from the star's envelope \cite{Kumar2008,Cannizzo2009,Cannizzo2011}, or the accretion disc enters a magnetically arrested disc stage, which allows the strong magnetic field to penetrate the black hole's ergosphere and extract its rotational kinetic energy through processes such as the \cite{BlandfordZnajek} mechanism, thus spinning it down in a manner similar to the magnetar model \cite{Lenart2025}. The diversity of possible central engines complicates the use of GRBs as cosmological probes. To account for this, we carefully selected a set of collapsar-origin GRBs likely powered by an ideal magnetar dipole-like central engine. In the ideal dipole spin-down approximation, the luminosity evolves as:
\begin{equation}
    L(t)=L_0 \left(1+\frac{t}{T_0}\right)^{-2}\,.
    \label{eq:magnetar}
\end{equation}
Here, $T_0$ is the characteristic spin-down timescale and $L_0=L(0)$ is the initial luminosity; the luminosity at $t=T_0$ is therefore $L(T_0)=L_0/4$.  This is the so-called ideal magnetar model. However, it is possible that the magnetic-field geometry of the neutron star is more complex than a simple dipole \cite{Wang2024}, and that radiative losses must also be accounted for. All these factors may contribute to the shape of the light curve, distorting the ideal picture, which in turn can introduce additional scatter into the empirical plateau luminosity-duration correlation. Therefore, \cite{Wang2022} concluded that one needs to select only GRBs with light curves closely following Eq. \ref{eq:magnetar}, in order to pinpoint a sample that tightly follows the aforementioned correlation. Indeed, for the ``ideal magnetar'' sample, the authors obtained very low scatter and were able to use it in cosmological analysis. However, the authors did not actually check whether ``ideal magnetar'' is the preferred model for the sources in their sample. Thus, in this work, we selected our sample based on the requirement that ``ideal magnetar'' must be the statistically preferred model compared to the empirical \cite{Willingale2007} model:
\begin{equation}
    F(t)=F_0e^{-\frac{t}{T}}\times \left\{\begin{array}{cc}
        e^{\alpha\left(1-\frac{t}{T}\right)} & t\leq T, \\
        \left(\frac{t}{T}\right)^{-\alpha} & t>T
    \end{array}\right.,
\end{equation}
where $F_0$ and $T$ are the flux at the end of the plateau and the observer-frame duration, respectively, and $\alpha$ is the slope of the light-curve decay after the plateau. We fitted both models to every source with a plateau and computed the Akaike Information Criterion (AIC). We then used it to calculate the relative probability that the magnetar model is preferred, given by:
\begin{gather}
        B_{\rm magnetar} = \exp\left(\frac{\rm min(AIC_{W07},AIC_{magnetar})-AIC_{magnetar}}{2}\right) \\
        B_{\rm W07} = \exp\left(\frac{\rm min(AIC_{W07},AIC_{magnetar})-AIC_{W07}}{2}\right) \\
        P_{\rm magnetar} = \frac{B_{\rm magnetar}}{B_{\rm magnetar}+B_{\rm W07}}.
\end{gather}
We selected only GRBs for which $P_{\rm magnetar}\geq 0.95$, thus ensuring a statistically significant sample. We applied this methodology to all 274 GRBs with an X-ray plateau and known redshift observed to date by the {\it Swift} satellite, and obtained a sample of 23 high-probability magnetar GRBs. We present the parameter space of the selected sample in Fig. \ref{fig:Magnetar_matrix}.

The simple MCMC fitting of the Dainotti relation with this sample yielded $a=-1.31  \pm 0.10$, $b=52.21  ^{+0.39}_{-0.41}$, $\sigma_{int}=0.34  ^{+0.04}_{-0.08}$. Thus, intrinsic scatter $\sigma_{\rm int}$ is reduced by 42\% compared to the platinum sample ($\sigma_{\rm int} = 0.59  ^{+0.05} _{-0.07}$). We emphasize here that the ideal magnetar model predicts the slope $a = -1$ \cite{Rowlinson:2014dja}, which is about 3$\sigma$ away from our estimate. However, correlations between the underlying magnetar parameters can modify the population slope. For fixed stellar structure, radiative efficiency, and beaming, dipole spin-down gives $L_a \propto B^2 P_0^{-4}$ and $T_a^* \propto P_0^2 B^{-2}$, where $B$ and $P_0$ denote the dipole magnetic field and initial spin period, respectively. An illustrative relation $B \propto P_0^q$ therefore yields $L_a \propto (T_a^*)^{(2q-4)/(2-2q)}$. For example, $q=-2.2$ gives $a=-1.3125$, consistent with our measured slope of $-1.31\pm0.10$. This illustrates how the observed slope can be accommodated within the dipole model, without establishing that this particular magnetic-field--spin-period anticorrelation is present in our sample.
However, it is worth noting that in \cite{Stratta:2018xza} a correlation between $P_0$ and $B$ is found as $P_0 \propto B^{6/7}$. If the initial period approximately follows the spin-up
equilibrium value, then
\begin{equation}
    P_0 \simeq P_{\rm eq}
    \propto B^{6/7}\dot{M}^{-3/7},
\end{equation}
where $\dot{M}$ is the accretion rate and the neutron-star mass and radius are held fixed \citep{Stratta2018}. Allowing $\dot M$ to vary changes the population prediction, $a_{\rm pop}=\frac{5-18u}{1+9u},\qquad u\equiv \frac{\operatorname{Var}(\log\dot M)}{\operatorname{Var}(\log B)}.$

Substituting the equilibrium-period scaling into the 
dipole luminosity relation yields:
\begin{equation}
    \begin{aligned}
        L_a
        &\propto B^2P_0^{-4}
        \propto B^2\left(B^{6/7}\dot{M}^{-3/7}\right)^{-4}\\
        &\propto B^{2-24/7}\dot{M}^{12/7}
        = B^{-10/7}\dot{M}^{12/7}.
    \end{aligned}
\end{equation}
Similarly, substituting into the plateau-duration relation gives
\begin{equation}
    \begin{aligned}
        T_a^*
        &\propto P_0^2B^{-2}
        \propto \left(B^{6/7}\dot{M}^{-3/7}\right)^2B^{-2}\\
        &\propto B^{12/7-2}\dot{M}^{-6/7}
        = B^{-2/7}\dot{M}^{-6/7}.
    \end{aligned}
\end{equation}
The negative magnetic-field exponents arise because $P_0$ varies with $B$ through the assumed equilibrium relation. In particular, the factor $P_0^{-4}$ contributes $B^{-24/7}$ to the luminosity, giving a net dependence of $B^{2-24/7}=B^{-10/7}$. These algebraic substitutions require $P_0\simeq P_{\rm eq}$, but do not require $B$ and $\dot{M}$ to be statistically uncorrelated. Let $b=\log B$ and $m=\log\dot{M}$, with population variances $\sigma_b^2$ and $\sigma_m^2$ and covariance $C_{bm}=\operatorname{Cov}(b,m)$. Denoting population means by overbars, define
\begin{align*}
    \sigma_b^2 &= \left\langle(b-\bar b)^2\right\rangle, \\
    \sigma_m^2 &= \left\langle(m-\bar m)^2\right\rangle, \\
    C_{bm} &= \left\langle(b-\bar b)(m-\bar m)\right\rangle.
\end{align*}
Thus, $\sigma_b^2$ and $\sigma_m^2$ measure the population spreads in the logarithms of magnetic field and accretion rate, while $C_{bm}$ describes their covariance. The population slope of the linear regression of $\log L_a$ on $\log T_a^*$ is therefore
\begin{equation}
    \begin{split}
    a_{\rm pop}
    &\equiv
    \frac{\operatorname{Cov}(\log L_a,\log T_a^*)}
         {\operatorname{Var}(\log T_a^*)}
    = \frac{5\sigma_b^2-18\sigma_m^2+9C_{bm}}
           {\sigma_b^2+9\sigma_m^2+6C_{bm}}.
    \end{split}
\end{equation}
The equilibrium relation does not, by itself, specify this population covariance. As an illustrative special case only, setting $C_{bm}=0$ and with equal logarithmic variances, $\sigma_m^2=\sigma_b^2$, correspond to $u=1$ and give
\begin{equation}
    a_{\rm pop}=\frac{5-18u}{1+9u},
    \qquad u=\frac{\sigma_m^2}{\sigma_b^2}.
\end{equation}
For $u\simeq1.0$, this yields $a_{\rm pop}\simeq-1.30$, consistent with our measured slope of $-1.31\pm0.10$. This example demonstrates mathematical compatibility with dipole spin-down and spin-up equilibrium under an assumed population distribution. The zero covariance and variance ratio used here are illustrative assumptions, rather than predictions of the magnetar model or measurements for our sample; reproducing the slope in this way does not constitute an independent test of the model. As the slope in our results is still within 3 $\sigma$ of the predicted -1, thus, we postpone further discussion in a near future once we reach a more sizable sample.

\subsection{New method for more reliable fits}

GRB light curves commonly suffer from gaps in the data or sparse sampling. Thus, when one fits raw light curve data to a given model through likelihood maximization, the maximum found might not correspond to the true physical solution. In fact, it is possible that several local maxima exist. This issue introduces a systematic source of scatter into the correlation. However, since it was established \cite{Zaninoni2011,Bernardini2012,Xu2012A&A...538A.134X,Mangano2012,Sultana_2012,Zaninoni2013,Margutti2013,Bardho2015,Izzo2015A&A...582A.115I,Kawakubo_2015,Si_2018,Zhao2019ApJ...883...97Z,Tang2019ApJS..245....1T,WangFeifei2020,Muccino2020,CaoShulei2022a,XuFan2021ApJ...920..135X,Levine2022,Shuang2022ApJ...924...69Y,Deng2023ApJ...943..126D,Tian_2023,Li_2023,Xu_2023,Deng2025} that the correlation exists \citep{Dainotti:2008vw}, is free from bias \citep{Dainotti2013a}, and is independent of evolutionary effects \citep{Dainotti2017a}, one can exploit the existence of the correlation to inform the likelihood estimator of the expected light-curve fit. Moreover, not to assume anything a priori, one can fit all the light curves from the GRB sample together with the correlation itself. In this scenario, the log-likelihood can be written as:
\begin{equation}
    \log(\mathcal{L}) = \sum_{i=1}^{N}\log(\mathcal{L}_{i\, {\rm LC}}(T_{a,i},F_{a,i})) +\gamma\log(\mathcal{L}_{\rm correlation}(T_{a,1}^*,...,T_{a,N}^*,L_{a,1},...,L_{a,N},a,b,\sigma_{\rm int})).
    \label{eq:new_like}
\end{equation}
We set $\gamma = 1$, so that each term in the joint likelihood enters
with its natural statistical weight.
$\mathcal{L}_{i\,\mathrm{LC}}(T_{a,i}, F_{a,i})$ is the likelihood
for the light-curve fit of the $i$-th GRB, which has $B_i$ data points:
\begin{equation}
\left\{\begin{array}{l}
    \mathcal{L}_{i\,\mathrm{LC}} =
        \prod_{j=1}^{B_i}
        \frac{1}{\sqrt{2\pi}\,\sigma_{j}}
        \exp\!\left[
            -\frac{\bigl(\log_{10}F_j
                   -\log_{10}F_{\mathrm{th}}(T_{a,i},F_{a,i},t_{j})
                   \bigr)^2}{2\sigma_{j}^2}
        \right],\\[10pt]
    \sigma_{j}=\sqrt{
        \left(
            \frac{\partial\log_{10}F_{\mathrm{th}}}{\partial t}
            \bigg|_{t=t_j}
            \sigma_{t_{j}}
        \right)^{\!2}
        + \sigma_{F_j}^2}.
\end{array}\right.
\end{equation}
$\mathcal{L}_{\mathrm{correlation}}$ is the likelihood of the plateau
duration--luminosity correlation:
\begin{equation}
    \mathcal{L}_{\mathrm{correlation}} =
        \prod_{i=1}^{N}
        \frac{1}{\sqrt{2\pi\,\tilde{\sigma}_i^2}}\,
        \exp\!\left[
            -\frac{\bigl(
                \log_{10}L_{a,i} - \log_{10}L_{\mathrm{th},i}
                \bigr)^2}{2\tilde{\sigma}_i^2}
        \right],
    \label{eq:cor_like}
\end{equation}
where $\log_{10}L_{\mathrm{th},i} = a\log_{10}T^*_{a,i} + b$ is the
correlation-predicted luminosity and
$\tilde{\sigma}_i^2 = \sigma_{\mathrm{int}}^2 + \sigma_{K,i}^2$
is the total variance per source, combining the intrinsic scatter
$\sigma_{\mathrm{int}}$ with the propagated $K$-correction uncertainty
$\sigma_{K,i}$.
The statistical uncertainties on $F_{a,i}$ and $T_{a,i}$ do not appear
separately in $\mathcal{L}_{\mathrm{correlation}}$: these quantities
are free parameters simultaneously constrained by
$\mathcal{L}_{i\,\mathrm{LC}}$ through the light-curve data, so their
uncertainty is already encoded in the posterior of the sampler.
The $K$-correction, by contrast, is not a free parameter and its
uncertainty does not enter $\mathcal{L}_{i\,\mathrm{LC}}$; it must
therefore appear explicitly in the correlation term.

We do not apply an additional weight (i.e.\ we set $\gamma=1$) for the
following reason. The global correlation parameters $a$, $b$, and
$\sigma_{\mathrm{int}}$ are constrained exclusively by
$\mathcal{L}_{\mathrm{correlation}}$, regardless of its numerical
magnitude relative to $\sum_i\mathcal{L}_{i\,\mathrm{LC}}$; no
weighting is needed for them to be well determined.
The only parameters that appear in both likelihood terms are the
cosmological parameters $H_0$ and $\Omega_m$, which enter through the
luminosity distance $d_L(z)$ used to convert observed flux to
luminosity.
Setting $\gamma=N$ would inflate the correlation term by a factor of
$N=23$ relative to both the light-curve term and the external
calibrators (CC, $N_{\mathrm{CC}}=32$ points; SNe~Ia,
$N_{\mathrm{SN}}=3714$ points), thereby artificially over-weighting
the GRB correlation in the inference of $H_0$ and $\Omega_m$.
We verified this empirically: with $\gamma=N$ the posteriors of $H_0$
and $\Omega_m$ exhibit anomalous shifts when cosmic chronometers are
included in the analysis, whereas $\gamma=1$ yields stable,
well-behaved posteriors across all dataset combinations considered in
this work.

Here, $T_{a,i}$ and $F_{a,i}$ are free parameters of the MCMC, simultaneously constrained by the light-curve likelihood $\mathcal{L}_{i\,\mathrm{LC}}$ through the light-curve data of the $i$-th GRB; they are not fixed to pre-computed best-fit values. The slope $a$, normalization $b$, and intrinsic scatter $\sigma_{\mathrm{int}}$ are the global parameters of the Dainotti correlation. The rest-frame plateau duration is $T^*_{a,i} = T_{a,i}/(1+z)$, and the plateau luminosity is $L_{a,i} = F_{a,i}\times 4\pi d_L^2(z)\times K(\beta,z)$, where $d_L(z)$ is the luminosity distance evaluated within the likelihood at the current cosmological parameters, and $K(\beta,z)$~\cite{Bloom2001} is the $K$-correction as a function of spectral index $\beta$ and redshift $z$. The statistical uncertainties on $T_{a,i}$ and $L_{a,i}$ do not appear explicitly in $\mathcal{L}_{\mathrm{correlation}}$ Eq.~\ref{eq:cor_like}) because these quantities are free parameters: their uncertainty is already encoded in the joint posterior through the light-curve term $\sum_i \mathcal{L}_{i\,\mathrm{LC}}$, which the sampler explores simultaneously with the correlation and cosmological parameters.

Applying this method to our magnetar data, we obtained: $a=-1.31\pm 0.11$ , $b=  52.20\pm 0.43 $, $\sigma_{\rm int}=   0.34\pm  0.06$. Thus, a reduction in intrinsic scatter, compared to the platinum sample, is a remarkable $42\%$. This analysis has been done with the Gaussian priors considering the cosmological $\Lambda$ CDM model with 5 $\sigma$ centered around $\Omega_M=0.3$ and $\sigma$ centered around $H_0=70$ km $s^{-1}$ $Mpc^{-1}$. We will use these units for $H_0$.

\begin{figure*}[t]
      \centering
      \includegraphics[width=0.99\linewidth]{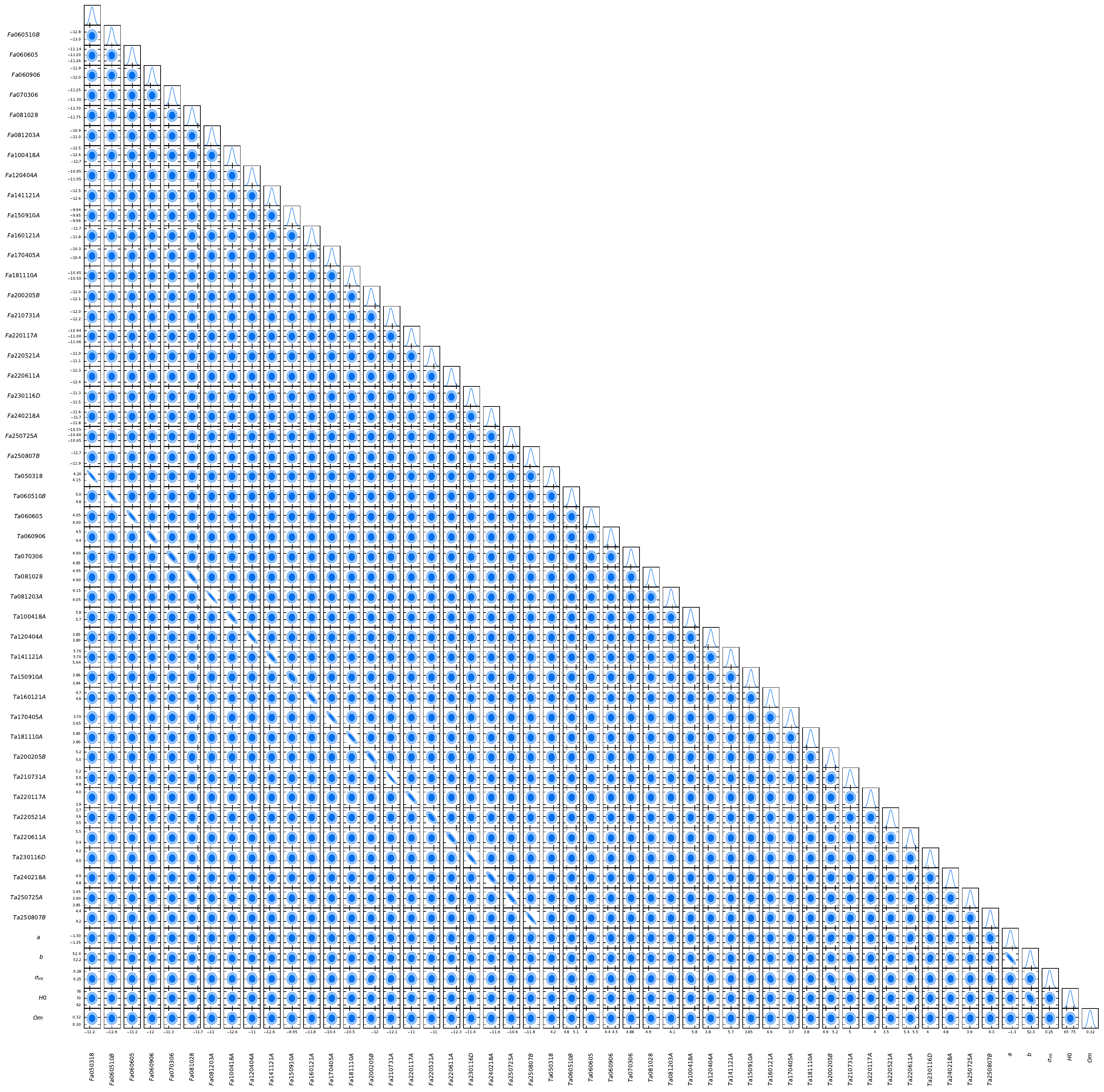}
      \caption{Posterior distributions from the joint light-curve and Dainotti correlation MCMC for the M23 sample, with Gaussian priors on $H_0$ and $\Omega_m$ centered on 70 and 0.3 respectively. Diagonal panels show marginalized 1D posteriors; off-diagonal panels show 2D credible regions at 68\% and 95\%. The 48 light-curve parameters ($F_{a,i},T_{a,i}$, $i = 1,\dots,23$) are fitted simultaneously with the Dainotti correlation parameters ($a,b,\sigma_{\rm int}$) and cosmological parameters ($H_0,\Omega_m$). This run serves to validate the joint likelihood and confirm that the correlation parameters are stable when the cosmological parameters are anchored externally.}
      \label{fig:GRB_calibration_with_SNe_Planck}
  \end{figure*}
This analysis uses Planck 2018 Gaussian priors on the cosmological parameters to validate the joint likelihood; the cosmology-unconstrained runs anchored to SNe~Ia and CC are presented in Section~\ref{sec:results}. Another challenge in GRB cosmology is the possible evolution of GRB properties with redshift. Indeed, in the general GRB population we observe that sources at higher redshift are more luminous and shorter-lived \cite{Lenart2025}. One possible explanation is that massive stars at high redshift may have had lower metallicity and could therefore have been more massive, producing brighter, more energetic bursts. From a theoretical point of view, \cite{Lenart2025} shows that, within both the magnetar and black hole models, evolution is expected to affect only the normalization of the correlation (i.e., the normalization becomes redshift dependent). However, our sample-selection method retains only likely ideal-magnetar GRBs powered by neutron stars of similar mass, so no significant evolution is expected. Still, for completeness we fitted the correlation while also accounting for redshift evolution. The standard procedure is to apply Kendall's $\tau$ statistic to determine the parameters of the function that best fits the evolutionary trend of a given physical variable \cite{Efron1992}. However, this requires relatively large samples, while we have a sample of only 23 sources. Thus, we treat the evolution correction as a free parameter of the correlation, by replacing $L_{a\, i}\to L_{a\, i}/(1+z)^{\kappa_{L_a}}$ and $T_{a\, i}^*\to T_{a\, i}^*/(1+z)^{\kappa_{T_a}}$ in Eq. \ref{eq:new_like}. These evolutionary parameters are degenerate in this type of analysis, so we impose Gaussian priors on them. With this procedure we obtained $a=-1.23 \pm 0.13$, $b=51.58 \pm 0.68$, $\sigma_{int}=0.35 \pm 0.06$, see Table III for the cosmological parameter results. Additionally, we performed a second computation in which the evolution is a simple additive term, substituting the correlation's normalization with the redshift-dependent form $b\to b+\kappa\times \log_{10} (1+z)$, obtaining $a=-1.24 \pm 0.13$, $b=51.68 \pm 0.74$, $\sigma_{int}=0.35 \pm 0.06$. The results of these two evolution treatments do not significantly change the best-fit estimates or uncertainties of either the light-curve or the correlation parameters; we therefore conclude that our sample is not significantly affected by evolution, and we can safely proceed without any correction. Theoretically, the evolution of the normalization of the correlation corresponds to the evolution of the initial kinetic energy of the magnetar, which is in agreement with expectations from stellar physics. The lower the metallicity, the higher the angular momentum of the star before its death; thus, a newborn magnetar can inherit a larger spin. 

\subsection{The traditional method}

The traditional GRB log-likelihood is
\begin{equation}
    \ln\mathcal{L}_{\rm GRB} = -\frac{1}{2}\sum_{i=1}^{N_{\rm GRB}}
    \left[\ln\sigma^2_i + 
    \frac{\left(\log L_{a,i}^{\rm obs} - \log L_{a,i}^{\rm th}
    \right)^2}{\sigma^2_i}\right]\,,
    \label{eq:grb_like}
\end{equation}
with total variance $\sigma^2_i = \sigma^2_{\rm int} + \sigma^2_{\log F_{a,i}} + a^2\,\sigma^2_{\log T_{a,i}}$, propagating flux and time uncertainties through the relation, where $a < 0$ is the slope, $b$ the intercept, and $\sigma_{\rm int}$ an intrinsic scatter term. We apply this method to the platinum dataset.

\subsection{Cosmological fits}

Equation \ref{eq:new_like} contains luminosities, which are distance-dependent. For the purpose of correlation studies, we calculate them assuming a cosmological model and values of its parameters. However, the cosmological parameters can be left free to vary within the likelihood-maximization algorithm, allowing us to estimate their values independently using GRBs.

Throughout this work, we adopt a spatially flat $\Lambda$CDM background cosmology, in which the dimensionless Hubble parameter takes the form
\begin{equation}
    E(z) \equiv \frac{H(z)}{H_0} = \sqrt{\Omega_m(1+z)^3 + (1 - \Omega_m)}\,,
\end{equation}
where $H_0$ is the Hubble constant and $\Omega_m$ is the present-day matter density parameter. The luminosity distance is
\begin{equation}
    d_L(z) = \frac{c(1+z)}{H_0}\int_0^z \frac{\mathrm{d}z'}{E(z')}\,,
\end{equation}
where $c$ is the speed of light. These two quantities enter the construction of the theoretical distance modulus for SNe~Ia and the theoretical plateau luminosity for GRBs, respectively, and constitute the cosmological model evaluated in all likelihood calculations.

Parameter inference is performed via Markov Chain Monte Carlo (MCMC) sampling using the \texttt{cobaya} package~\cite{Torrado:2020dgo}, with convergence assessed through the Gelman--Rubin criterion, requiring $R - 1 < 0.01$ for all chains. Posterior distributions are analyzed and visualized using \texttt{GetDist}~\cite{Lewis:2019xzd}. All priors are specified individually for each analysis and are detailed in Section~\ref{sec:results}.

We began with a baseline cosmological estimation from the SNe~Ia master sample alone, which serves as both a pipeline validation and a reference point common to both GRB  analyses. We then perform two independent calibrations of the GRB correlation parameters, one anchored to SNe~Ia and one to Cosmic Chronometers, before concluding with the full cosmological inference under progressively richer probe combinations.

The calibration with CC was performed with 23 GRBs up to redshift $z<1.965$; the obtained fitting parameters with the new method are: $a = -1.32 \pm 0.10$, $b = 52.29 \pm 0.41$, $\sigma_{\rm int} = 0.35 \pm 0.06$, for the magnetar sample and $a = -0.94 \pm 0.16$, $b = 52.03 \pm 0.53$, $\sigma_{\rm int} = 0.59 ^{+0.05}_{-0.07}$, for the platinum sample. While the calibration with SNe was performed with 23 GRBs up to redshift $z<2.26$; the obtained fitting parameters are: $a = -1.33 \pm 0.02$, $b = 52.31 \pm 0.07$, $\sigma_{\rm int} = 0.26 \pm 0.01$, for the magnetar sample and $a = -0.93 ^{+0.17}_{-0.16}$, $b = 51.98 ^{+0.52} _{-0.54}$, $\sigma_{\rm int} = 0.59 ^{+0.05} _{-0.07}$, for the platinum sample. Further, the obtained parameters and their uncertainties were used as Gaussian priors in the cosmological fitting of full GRB samples.

\section{Results}
\label{sec:results}

We present our results in three stages of increasing complexity, performing every analysis independently for the two GRB samples (magnetar and platinum)  and comparing the outcomes at each stage. This parallel treatment allows us to assess the robustness of the cosmological inference against the choice of GRB  selection criteria, since the two samples differ significantly in size ($23$  vs $50$ GRBs) and in the way they are built. 

\subsection{Baseline: SNe~Ia constraints}
\label{sec:sn_only}

As a first step, we estimate the cosmological parameters from the SNe~Ia master sample alone. The dataset comprises $3714$ data points spanning $z \in [0.00122,\, 2.261]$. The theoretical distance modulus is
\begin{equation}
    \mu_{\rm th}(z) = 5\log_{10}\!\left[d_L(z, H_0, \Omega_m)\right] + 25\,,
\end{equation}
where $d_L$ is the luminosity distance evaluated in a flat $\Lambda$CDM background with $E(z) = \sqrt{\Omega_m(1+z)^3 + (1-\Omega_m)}$. The likelihood is
\begin{equation}
    \ln\mathcal{L}_{\rm SN} = 
    -\frac{1}{2}\,\Delta\boldsymbol{\mu}^{\top}\,
    \mathbf{C}^{-1}_{\rm SN}\,\Delta\boldsymbol{\mu}\,,
\end{equation}
where $\Delta\boldsymbol{\mu} = \boldsymbol{\mu}_{\rm obs} - \boldsymbol{\mu}_{\rm th}$ and $\mathbf{C}_{\rm SN}$ is the full covariance matrix accounting for statistical and systematic uncertainties. Uniform priors $H_0 \in [60, 80]\ \rm km\,s^{-1}\,Mpc^{-1}$ and $\Omega_m \in [0.1, 0.5]$ are adopted. This analysis is independent of the GRB sample choice and serves as the common cosmological anchor for all subsequent runs.

\begin{figure}[t]
      \centering
      \includegraphics[width=0.5\linewidth]{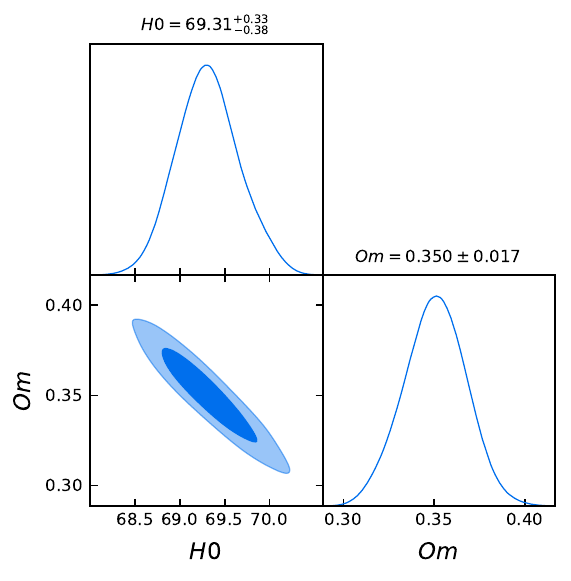}
      \caption{Marginalized posterior distributions of $H_0$ and $\Omega_m$ from
  the SNe~Ia master sample alone ($3714$ data points). Contours show the $68\%$
  and $95\%$ confidence levels.}
      \label{fig:H0-Om_master_sample}
  \end{figure}

The SNe~Ia-only analysis yields
\begin{equation*}
    H_0 = 69.3 \pm 0.36\ \rm km\,s^{-1}\,Mpc^{-1}\,, \qquad 
    \Omega_m = 0.35 \pm  0.02\,,
\end{equation*}
at $68\%$ confidence level, shown in Fig.~\ref{fig:H0-Om_master_sample}. The $H_0$ value sits between the Planck CMB estimate~\cite{Planck:2018vyg} and the SH0ES measurement~\cite{Riess:2021jrx}, as expected for a dataset that does not include the SH0ES Cepheid calibration of $M_B$. These constraints serve as the cosmological anchor for the GRB calibration that follows.

\subsection{Calibration of the GRB correlation parameters}
\label{sec:grb_calibration}

The observed plateau luminosity is computed from the measured flux $F_a$ as
\begin{equation}
    \log L_a^{\rm obs} = \log F_a + 2\log d_L + \log(4\pi) + K_a\,,
\end{equation}
where $d_L$ is evaluated at the GRB redshift under flat $\Lambda$CDM and $K_a$ is the $K$-correction provided in the dataset. This likelihood is evaluated identically for both GRB samples; the only difference between the two analyses is the dataset passed to it, and the likelihood used as discussed in Sec. \ref{sec:method}.

We calibrate $(a, b, \sigma_{\rm int})$ via two independent routes for each GRB sample, yielding four calibration runs in total.

\paragraph{Calibration with SNe~Ia.}
The joint likelihood $\ln\mathcal{L}_{\rm SN} + \ln\mathcal{L}_{\rm GRB}$ is   sampled over the full parameter space $(H_0, \Omega_m, a, b, \sigma_{\rm int})$ with uniform priors $H_0 \in [60, 80]\ \rm km\,s^{-1}\,Mpc^{-1}$, $\Omega_m \in [0.1, 0.5]$, $a \in [-3, 0]$, $b \in [40, 60]$, $\sigma_{\rm int} \in [0.01, 1.0]$. The SNe~Ia anchor the distance scale, allowing the GRB parameters to be calibrated without assuming a fiducial cosmology. Results for both samples are shown in Fig.~\ref{fig:GRB_calibration_with_SNe_Planck}. It is worth noting that for the magnetar sample, the light curve parameters are still a part of the fitting, but we do not display them for clarity.

\paragraph{Calibration with Cosmic Chronometers.}
The CC likelihood follows the equation
\begin{equation}
      \ln\mathcal{L}_{\rm CC} = -\frac{1}{2}\sum_{i=1}^{32}
      \left(\frac{H_{\rm obs}(z_i) - H_0\,E(z_i)}{\sigma_{H,i}}
      \right)^2\,,
      \label{eq:cc_like}
\end{equation}
adopting a diagonal covariance following standard practice~\cite{Yu:2017iju,  Gomez-Valent:2018hwc}, with 32 measurements spanning $z \in [0.07,1.965]$. The joint likelihood $\ln\mathcal{L}_{\rm CC} +   \ln\mathcal{L}_{\rm GRB}$ is sampled with the same uniform priors for both GRB samples, yielding the results shown in Fig.~\ref{fig:GRB_calibration_with_CC}.

\paragraph{Comparison.}
The calibrated Dainotti parameters from all six runs, i.e. the previous four together with the old and new method established here, are summarized in  Table~\ref{tab:calibration}. The most striking feature of this comparison is the intrinsic scatter $\sigma_{\rm int}$: although the magnetar sample contains less than half the number of GRBs of the platinum sample ($23$ vs $50$), it yields $\sigma_{\rm int} \approx 0.35$ for the CC-anchored run and a striking $\sigma_{\rm int} \approx 0.26$ for the SNe~Ia-anchored run, compared to $\sigma_{\rm int} \approx 0.59$ for platinum, a reduction of $\sim 41\%$ and $\sim 56 \%$ respectively. This indicates that the magnetar spin-down selection isolates a physically more homogeneous population than the platinum quality cuts alone can achieve and allows a reduction on the uncertainties on the cosmological parameters. The slope $a$ also differs substantially between samples ($a \approx -1.3$ for magnetar versus $a \approx -0.93$ for platinum), and this shift is reproduced identically by both calibrators, confirming that it reflects a genuine difference between the two GRB selections rather than a calibration artifact. By contrast, the cosmological parameters $H_0$ and $\Omega_m$ are driven primarily by the SNe~Ia or CC anchor rather than by the GRB sample, so their consistency across the two samples is expected and serves as a further validation of the pipeline.

Comparing the two analysis strategies applied to the magnetar sample, the standard Dainotti likelihood and the joint light-curve and correlation likelihood yield fully consistent calibration parameters across both anchors. The slope $a$, intercept $b$, and intrinsic scatter $\sigma _{\rm int}$ are indistinguishable between M23 (old) and M23 (new) to within the quoted uncertainties, confirming that the joint method does not bias the Dainotti parameters. The primary advantage of the joint likelihood therefore lies not in shifting the calibrated parameter values but in the reliability of the individual plateau fits: by simultaneously constraining $F_{a,i}$ and $T_{a,i}$ through both the light-curve data and the correlation, the method avoids the local maxima that can arise when fitting sparse or gapped light curves independently, producing a more physically grounded posterior for each GRB even when the marginalized Dainotti parameters remain unchanged.

\begin{table}[t]
  \centering
  \caption{Calibrated Dainotti and cosmological parameters from all calibration runs. All values are marginalized means with $68\%$ confidence intervals. M23 and P50 refer to the magnetar and platinum GRB samples respectively. M23 (old) uses the traditional Dainotti likelihood of Eq.~\eqref{eq:grb_like}; M23 (new) uses the novel joint light-curve and correlation likelihood of Eq.~\eqref{eq:new_like}.}
  \label{tab:calibration}
  \begin{tabular}{lcccccc}
  \hline\hline
   & \multicolumn{3}{c}{SNe~Ia $+$ GRB}
   & \multicolumn{3}{c}{CC $+$ GRB} \\
  \cmidrule(lr){2-4}\cmidrule(lr){5-7}
  Parameter & M23 (old) & M23 (new) & P50 & M23 (old) & M23 (new) & P50 \\
  \hline
  $H_0\ [\rm km\,s^{-1}\,Mpc^{-1}]$ & $69.30^{+0.35}_{-0.34}$ & $69.08 \pm 0.34$ & $69.32 ^{+0.36}_{-0.35}$
    & $67.61  ^{+2.87}_{-3.12}$ & $67.44 \pm 3.08$ & $67.78 ^{+2.96}_{-2.91}$ \\
  $\Omega_m$ & $0.35 \pm 0.02$
      & $0.36 \pm 0.02$ & $0.35 \pm 0.02$
      & $0.33^{+0.05}_{-0.06}$ & $0.34 \pm 0.06$ & $0.33^{+0.05}_{-0.06}$ \\
  $a$
    &  $-1.32 ^{ +0.11} _{ -0.10}$ & $-1.33 \pm 0.02$ & $-0.93 ^{+0.17}_{-0.16}$
    & $-1.32 \pm 0.10$ & $-1.32 \pm 0.10$ & $-0.94 \pm 0.16$ \\
  $b$
    & $52.24  \pm 0.41$ & $52.31 \pm 0.07$ & $51.98 ^{+0.52}_{-0.54}$
    & $52.30 \pm 0.41$ & $52.29 \pm 0.41$ & $52.03 \pm 0.53$ \\
  $\sigma_{\rm int}$
    & $0.35  ^{+0.04} _{-0.06}$ & $0.26 \pm 0.01$ & $0.59^{+0.05}_{-0.07}$
    & $0.35 ^{+0.04}_{-0.07}$ & $0.35 \pm 0.06$ & $0.59^{+0.05}_{-0.07}$ \\
  \hline\hline
  \end{tabular}
\end{table}

 \begin{figure*}[t]
      \centering
      \includegraphics[width=0.49\linewidth]{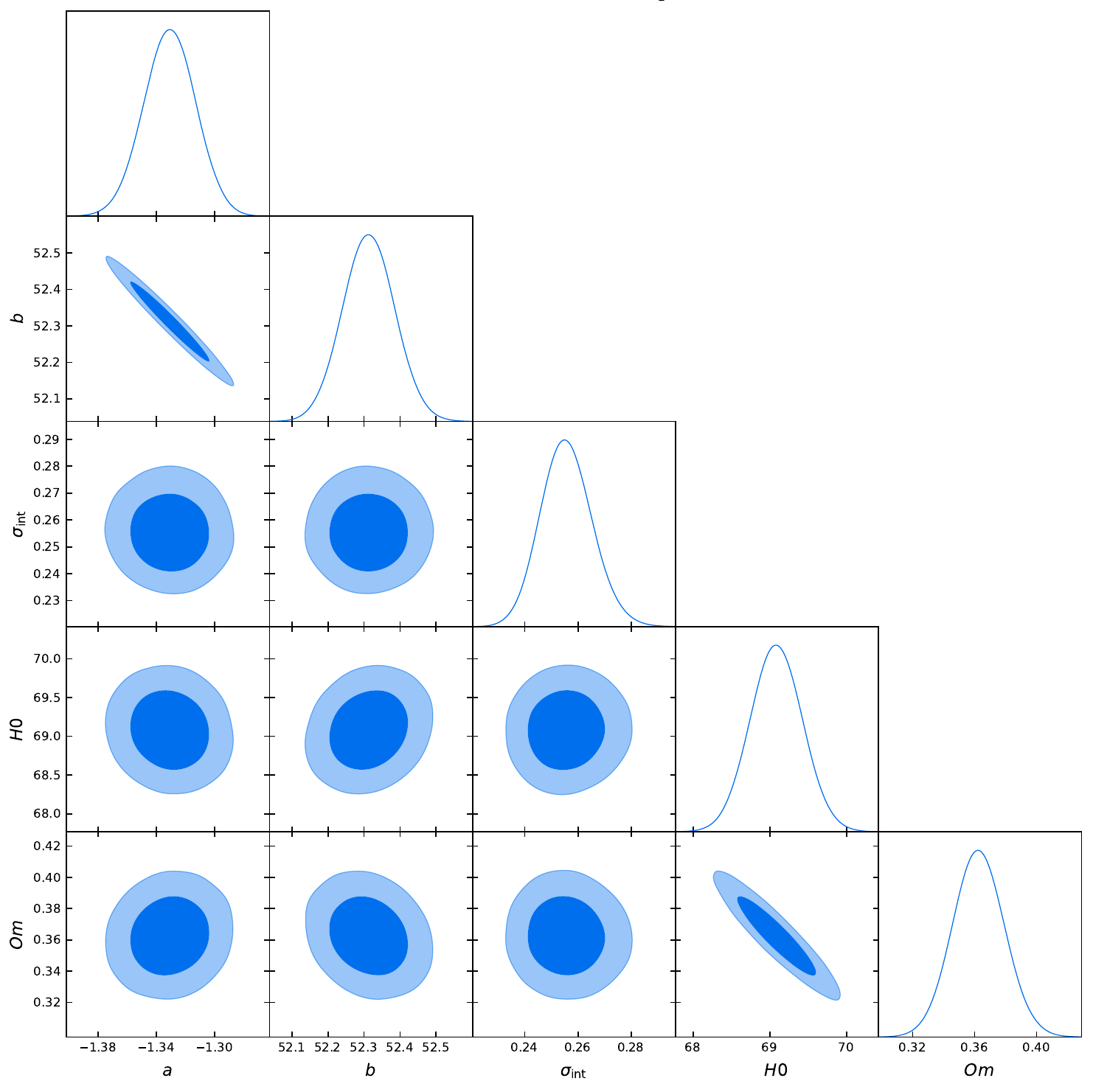}
      \includegraphics[width=0.49\linewidth]{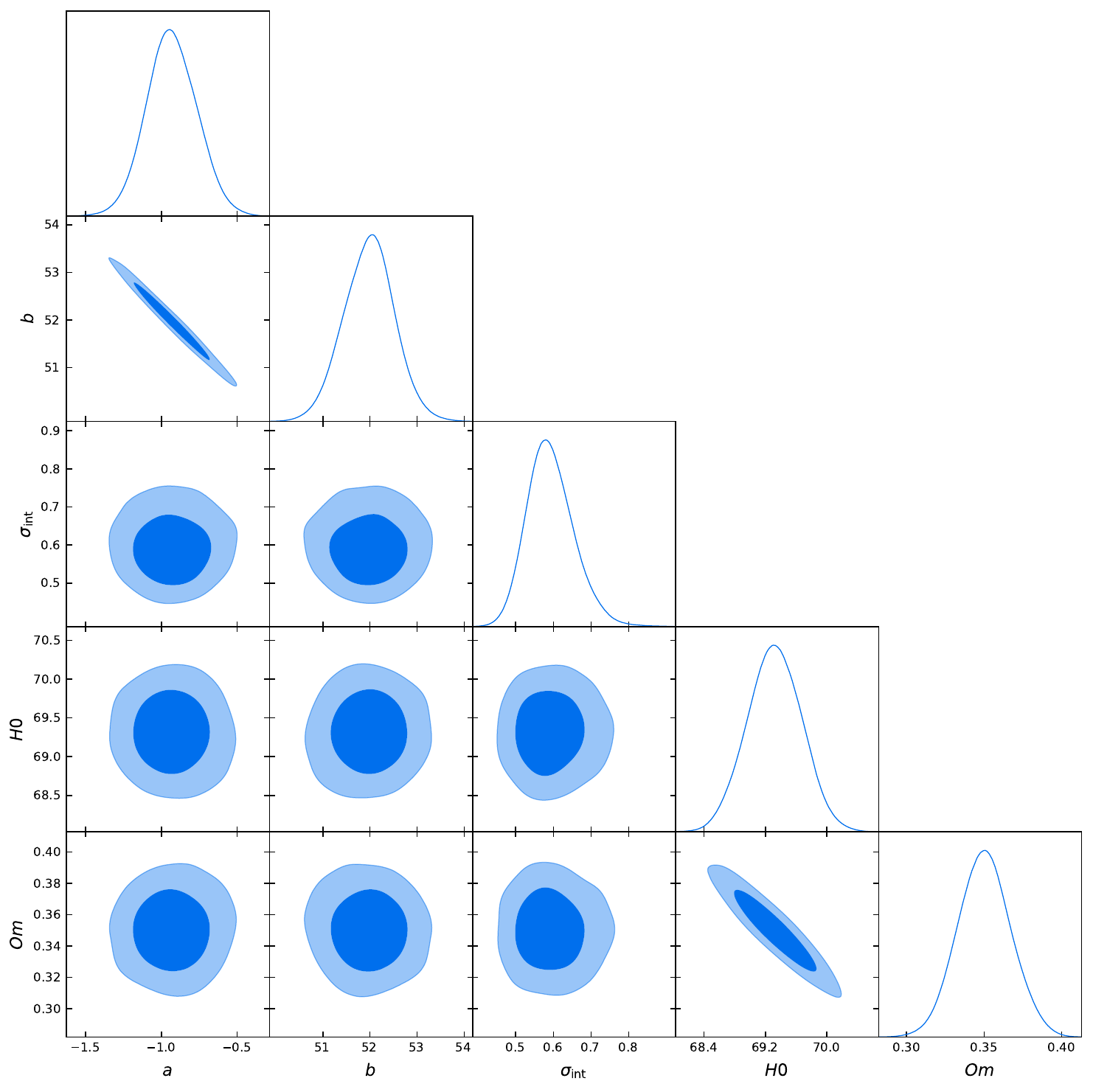}
      \caption{Joint SNe~Ia$+$GRB calibration of the Dainotti parameters $(a, b, \sigma_{\rm int},H_0, \Omega_m)$ for the magnetar (left) and platinum (right) samples. Contours show the $68\%$ and $95\%$ confidence levels. For the M23 sample the joint light-curve and correlation likelihood of Eq.~\eqref{eq:new_like} is used. The calibrated cosmological parameters are consistent between the two samples within $1\sigma$.}
      \label{fig:GRB_calibration_with_SNe}
  \end{figure*}

  \begin{figure*}[t]
      \centering
      \includegraphics[width=0.49\linewidth]{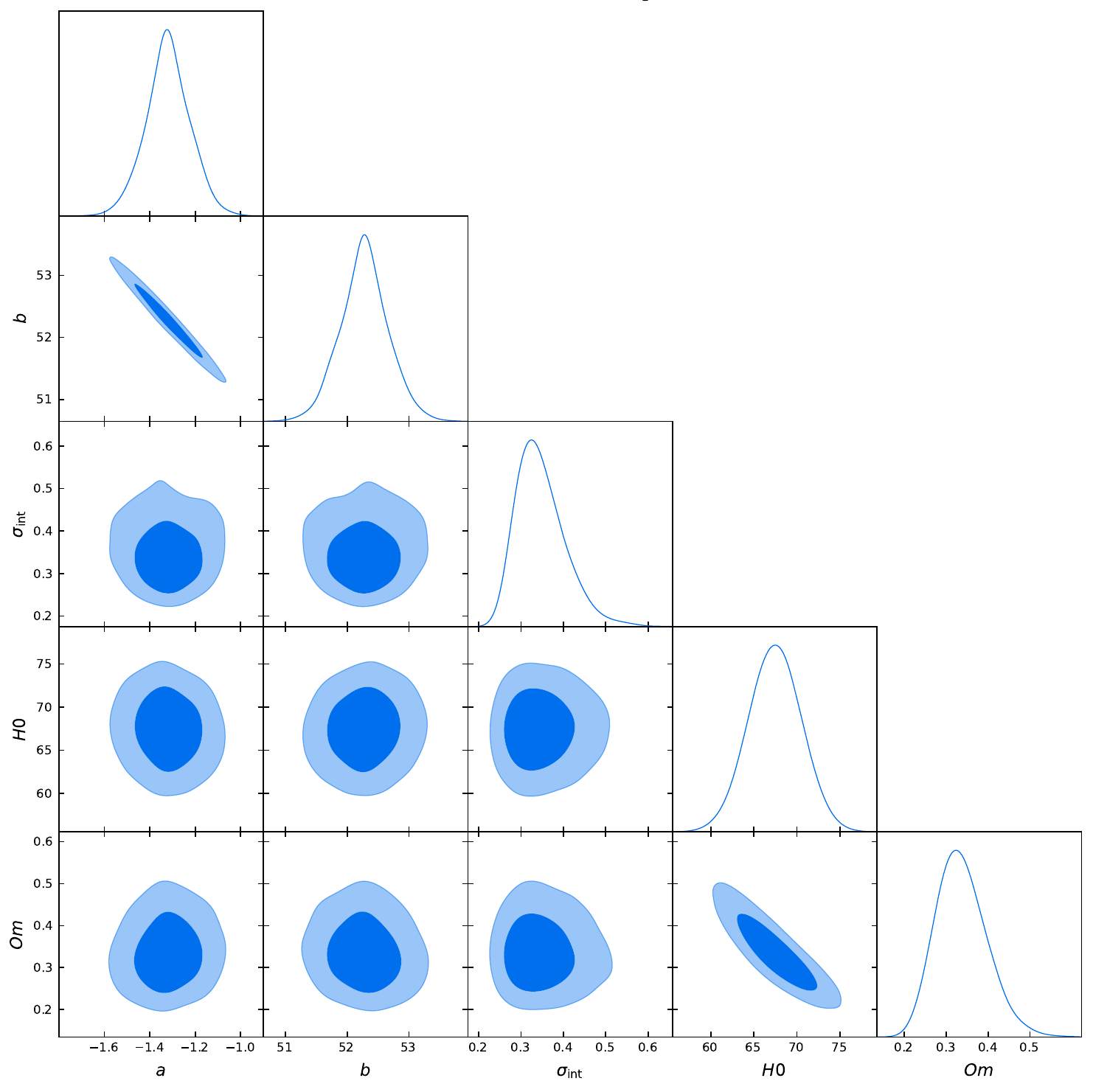}
      \includegraphics[width=0.49\linewidth]{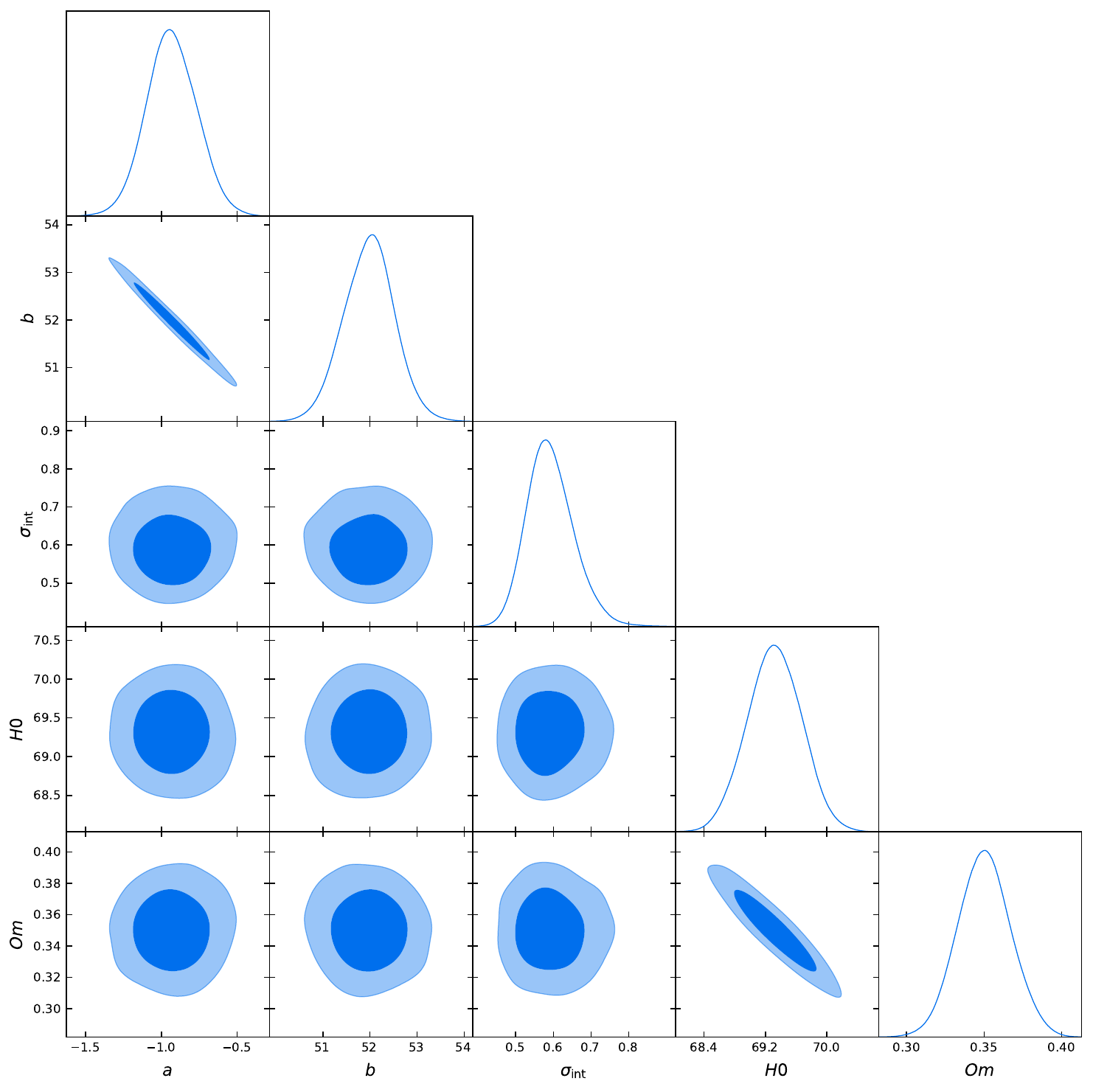}
      \caption{Joint CC$+$GRB calibration of the Dainotti parameters $(a, b, \sigma_{\rm int},H_0, \Omega_m)$ for the magnetar (left) and platinum (right) samples. Contours show the $68\%$ and $95\%$ confidence levels. For the M23 sample, the joint light-curve and correlation likelihood of Eq.~\eqref{eq:new_like} is used; for P50 the traditional likelihood of Eq.~\eqref{eq:grb_like} is used. As in Fig.~\ref{fig:GRB_calibration_with_SNe}, the cosmological parameters agree between samples while the Dainotti slope and scatter do not.}
      \label{fig:GRB_calibration_with_CC}
  \end{figure*}

\subsection{Cosmological parameter estimation}
\label{sec:cosmo_estimation}

\begin{figure}[t]
      \centering
      \includegraphics[width=0.7\linewidth,trim={0 0 0 2mm},clip]{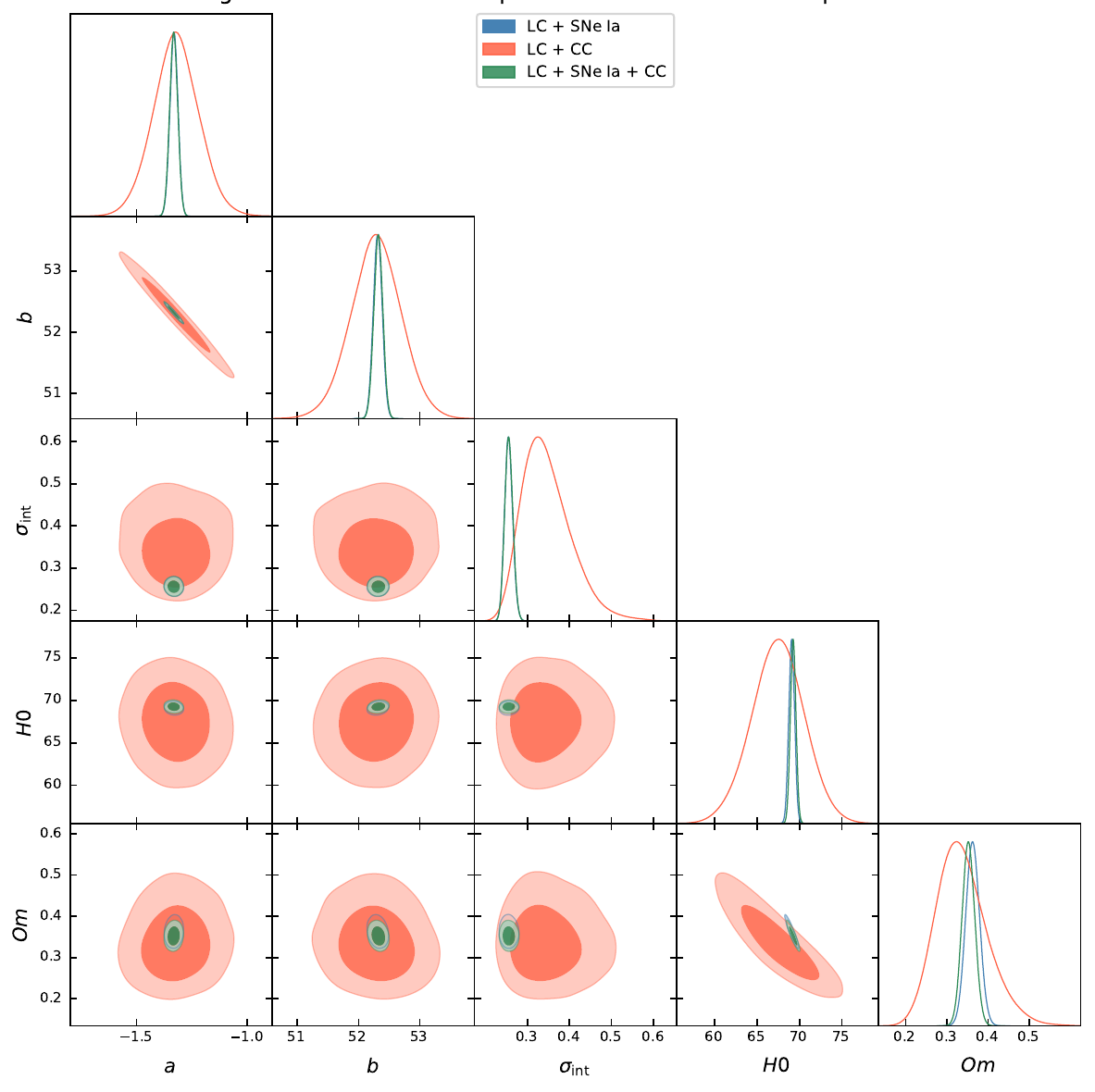}
      \caption{Comparison of the marginalized $H_0$--$\Omega_m$ posteriors from all probe combinations for the M23 sample, using the joint light-curve and correlation likelihood of Eq.~\eqref{eq:new_like}. Contours correspond to GRB+CC (orange contours), SNe~Ia+GRB (blue), and the full SNe~Ia+CC+GRB joint analysis (green), all at 68\% and 95\% confidence. The progressive improvement in constraining power from the addition of each probe is clearly visible.}
      \label{fig:comparison_M24}
  \end{figure}

\begin{figure}[t]
      \centering
      \includegraphics[width=0.7\linewidth]{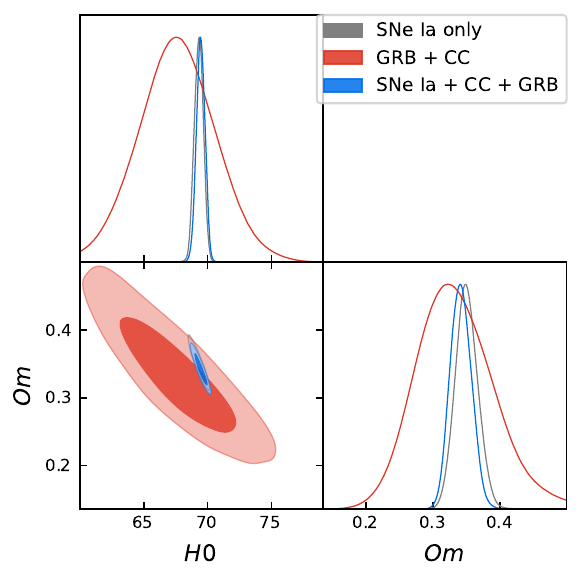}
      \caption{Comparison of the marginalized $H_0$--$\Omega_m$ posteriors from all probe combinations for the platinum sample. The traditional likelihood of Eq.~\eqref{eq:grb_like} is used throughout.}
      \label{fig:comparison_P50}
  \end{figure}

We now use the calibrated GRB parameters to constrain cosmology under three probe combinations of increasing constraining power, performing each analysis independently for both GRB samples. In all cases, the Dainotti nuisance parameters carry the information propagated from the calibration step: when GRBs appear alone or combined with CC, $(a, b, \sigma_{\rm int})$ are assigned Gaussian priors from the respective SNe~Ia calibration posteriors; when SNe~Ia are included directly in the likelihood, the nuisance parameters are left free with uniform priors to avoid double-counting. The cosmological results are summarized in Table~\ref{tab:cosmo_results} and the marginalized posteriors are compared in Figs.~\ref{fig:comparison_M24} and~\ref{fig:comparison_P50}.

GRBs alone can simultaneously constrain $H_0$ and $\Omega_m$, but they carry large uncertainties due to the dominance of intrinsic scatter $\sigma_{\rm int}$ over the  $d_L(z)$\citep{Dainotti2023alternative,Dainotti2023MNRAS.518.2201D}; we therefore proceed directly to the combinations with external anchors.

\paragraph{GRB $+$ CC.}
Adding the CC likelihood of Eq.~\eqref{eq:cc_like} breaks the $H_0$--$\Omega_m$ degeneracy for both samples. For magnetar, $H_0 =67.23^{+3.12}_{-3.17}\ \rm km\,s^{-1}\,Mpc^{-1}$ and $\Omega_m = 0.34 ^{+0.05}_{-0.07}$; for platinum, $H_0 = 67.8 \pm 2.9\ \rm km\,s^{-1}\,Mpc^{-1}$ and $\Omega_m = 0.33^{+0.05}_{-0.06}$. The GRBs extend the redshift lever arm well beyond the CC sample ($z \lesssim 2$), providing sensitivity to the distance--redshift relation at epochs unreachable by chronometers alone. The two samples yield fully consistent constraints on $\Omega_m$, with comparable posterior widths, confirming that the breaking of the $H_0$--$\Omega_m$ degeneracy is driven by the CC data rather than by the properties of the GRB sample.

\paragraph{SNe~Ia $+$ CC $+$ GRB.}
In the full joint analysis, all three likelihoods are combined simultaneously and $(a, b, \sigma_{\rm int})$ are sampled freely alongside $(H_0, \Omega_m)$. For magnetar, this yields $H_0 = 69.45 \pm 0.32\ \rm km\,s^{-1}\,Mpc^{-1}$ and $\Omega_m = 0.34 ^{+0.01}_{-0.02}$; for platinum, $H_0 = 69.46 \pm 0.33\ \rm km\,s^{-1}\,Mpc^{-1}$ and $\Omega_m = 0.34^{+0.015}_{-0.02}$. In both cases the results are fully consistent with the SNe~Ia-only baseline at the $1\sigma$ level, confirming that the GRB and CC data do not introduce systematic biases into the inference. For magnetar, $H_0$ deviates from the Planck CMB estimate of $67.4 \pm 0.5\ \rm km\,s^{-1}\,Mpc^{-1}$~\cite{Planck:2018vyg} at $3.45\sigma$, and from the SH0ES measurement of $73.04 \pm 1.04\ \rm km\,s^{-1}\,Mpc^{-1}$~\cite{Riess:2021jrx} at $3.27\sigma$. For platinum, the corresponding tensions with Planck and SH0ES are $3.43\sigma$ and $3.28\sigma$ respectively, in close agreement with the magnetar result (within $0.04\sigma$ in both cases). The agreement between the two samples in the joint run is remarkable:$H_0$ and $\Omega_m$ are compatible within 1 $\sigma$ to the precision reported, confirming that the cosmological inference is driven by the SNe~Ia and CC anchors.

\begin{table}[t]
\centering
\caption{Marginalized cosmological parameter constraints at $68\%$ confidence from all probe combinations for both GRB samples. M23 and P50 refer to magnetar and platinum respectively.}
  \label{tab:cosmo_results}
  \resizebox{0.7\columnwidth}{!}{%
  \begin{tabular}{lcccc}
  \hline\hline
   & \multicolumn{2}{c}{$H_0\ [\rm km\,s^{-1}\,Mpc^{-1}]$}
   & \multicolumn{2}{c}{$\Omega_m$} \\
  \cmidrule(lr){2-3}\cmidrule(lr){4-5}
  Dataset & M23 & P50 & M23 & P50 \\
  \hline
  SNe~Ia only
      & \multicolumn{2}{c}{$69.3 \pm 0.35$}
      & \multicolumn{2}{c}{$0.35 \pm 0.02$} \\
  GRB $+$ CC
      & $67.23^{+3.12}_{-3.17}$   & $67.58 ^{+3.02}_{-2.95}$
      & $0.34 ^{+0.05}_{-0.07}$ & $0.33^{+0.05}_{-0.06}$ \\
  SNe~Ia $+$ CC $+$ GRB
      & $69.45 \pm 0.32$ & $69.46 \pm 0.33$
      & $0.34 ^{+0.01}_{-0.02}$ & $0.34\pm 0.02$ \\
  \hline\hline
  \end{tabular}}
\end{table}

\begin{table}[t]
\centering
\caption{Marginalized cosmological parameter constraints at $68\%$ confidence from the M23 sample under two redshift-evolution corrections to the Dainotti relation. Fixed-$\kappa$: Efron--Petrosian Gaussian priors $\kappa_{L_a}=2.42\pm0.58$, $\kappa_{T_a}=-1.25\pm0.28$. Free-$\kappa$: evolution absorbed via $b\to b+\kappa\log_{10}(1+z)$ with $\kappa$ free. Baseline (no evolution) values from Table~\ref{tab:cosmo_results} are shown for comparison.}
\label{tab:evolution_results}
\resizebox{0.6\columnwidth}{!}{%
\begin{tabular}{lcc}
\hline\hline
Dataset & $H_0\ [\rm km\,s^{-1}\,Mpc^{-1}]$ & $\Omega_m$ \\
\hline
\multicolumn{3}{l}{\textit{Baseline (no evolution correction, M23)}} \\
GRB $+$ CC          & $67.23^{+3.12}_{-3.17}$    & $0.34^{+0.05}_{-0.07}$ \\
SNe~Ia $+$ CC $+$ GRB & $69.45\pm0.32$ & $0.34 ^{+0.01}_{-0.02}$ \\
\hline
\multicolumn{3}{l}{\textit{Fixed-$\kappa$ evolution}} \\
GRB $+$ CC          & $67.79 \pm 3.08
 $   & $0.33 \pm 0.06$ \\
SNe~Ia $+$ CC $+$ GRB & $69.47 \pm 0.33$
  & $ 0.34 \pm 0.02$ \\
\hline
\multicolumn{3}{l}{\textit{Free-$\kappa$ evolution}} \\
GRB $+$ CC          & $67.76 \pm 3.10$ & $0.33 \pm 0.06$ \\
SNe~Ia $+$ CC $+$ GRB & $69.47 \pm 0.33$ &
  $0.34 \pm 0.02$ \\
\hline\hline
\end{tabular}}
\end{table}

\section{Discussion and Conclusions}
\label{sec:conclusions}

In this work, we have investigated the cosmological constraining power of GRB plateau emission standardized through the two-dimensional Dainotti relation, using two samples that differ fundamentally in their selection philosophy. The magnetar sample (M23, 23 GRBs) is constructed by requiring that the ideal magnetar spin-down model is statistically preferred over the empirical Willingale parametrization, while the platinum sample (P50, 50 GRBs) applies the most stringent phenomenological quality cuts. Although the cut on the GRB sample is comparable to that of SNe Ia ($\sim 70\%$), the high-quality precision data are more challenging to obtain compared to other probes. This is the first time we have used the SNe Ia Master sample in which a new likelihood analysis has been performed following the approach in \cite{Dainotti2024JHEAp..41...30D} in combination with the 2D Danotti relation corrected for selection biases and redshift evolution. For M23 we introduced a novel joint likelihood that simultaneously fits the individual plateau light curves and the Dainotti correlation, allowing a better regularization of the light-curve fits and avoiding local maxima that arise from sparse or gapped data. For P50 we applied the standard Dainotti likelihood with data that are more well-sampled compared to the sample based on the magnetar. Both samples were calibrated independently against the Master SNe~Ia compilation \citep{Dainotti2025JHEAp..4800405D} and against Cosmic Chronometers, and the cosmological inference was performed in a spatially flat $\Lambda$CDM background sampled via \texttt{cobaya}. The inferred slopes remain consistent within their quoted uncertainties across the baseline and redshift-evolution treatments, indicating that these corrections do not substantially alter the measured luminosity--duration slope. In addition, the calibration of GRBs either with the Master sample or with the CC does carry the same stable parameter values both for the correlation and cosmology.

Although the calibration method applied by us is sufficient for our purposes given the small sample size, we note here that different methods exist. \cite{Shah:2024slr} created a Neural Network model \texttt{LADDER}, which was demonstrated to work well with GRB data, and allows GRBs to constrain cosmological parameters with similar precision as other calibration methods. In another work, \cite{Ruchika2026} demonstrated how to properly use Gaussian Process reconstruction, in a way that removes the systematics related to the choice of the kernel. Both these methods can be applied to our samples, although for the magnetar sample to work properly, we likely need more data.

The central result of this analysis is the intrinsic scatter of the Dainotti relation: the magnetar selection yields $\sigma_{\rm int} = 0.35$ for the CC-anchored runs and $\sigma_{\rm int} = 0.26$ for the SNe~Ia-anchored runs, compared to $\sigma_{\rm int} = 0.59$ for the platinum sample, a reduction of $\sim 41\%$ and $\sim 56\%$ respectively. This demonstrates that a physically motivated selection criterion, which retains only GRBs whose light curves are consistent with ideal dipole magnetar spin-down, removes a source of physical dispersion that generic quality cuts alone cannot eliminate. Compared with the broader GRB population, the M23 sample exhibits plateau duration uncertainties that are a factor of two smaller and luminosity uncertainties that are $1.5$ times smaller (in terms of median uncertainty), confirming the statistical advantage of this selection. For the cosmological inference, the full SNe~Ia$+$CC$+$GRB 
joint analysis yields $H_0 = 69.45 \pm 0.32\ \mathrm{km\,s^{-1}\,Mpc^{-1}}$ 
and $\Omega_m = 0.34 ^{+0.01}_{-0.02}$ for the magnetar sample, in $3.45\sigma$ 
tension with Planck~\cite{Planck:2018vyg} and $3.27\sigma$ tension with 
SH0ES~\cite{Riess:2021jrx}. The cosmological constraints are primarily driven by the SNe Ia and CC anchors. The advantage of GRBs is that they are observed at much larger redshifts than the SNe Ia, for example at $z=5.0$ for the platinum and $z=5.57$ for the magnetar sample. This high-redshift reach makes GRBs uniquely suited to the open question of cosmic tensions.

Comparing the two analysis strategies applied to M23, the traditional Dainotti likelihood and the joint light-curve and correlation likelihood yield fully consistent calibration parameters across both anchors. The slope $a$, intercept $b$, and intrinsic scatter $\sigma_{\rm int}$ are indistinguishable between the two methods to within quoted uncertainties. The advantage of the joint method therefore lies not in shifting the marginalized Dainotti parameters but in the reliability of the individual plateau fits: by simultaneously constraining $F_{a,i}$ and $T_{a,i}$ through both the light-curve data and the correlation, the method produces physically grounded posteriors for each GRB even when the marginalized parameters remain unchanged. Moreover, the new method, through its construction, takes into account the possible correlation between the uncertainty on both plateau flux and duration. We further tested the robustness of the results against redshift evolution of the Dainotti relation,  treating the evolution index both as a fixed parameter derived through the Efron--Petrosian method \cite{EfronPetrosian1992} and as a free parameter. In both cases, the cosmological constraints shift by less than $0.1\sigma$ relative to the no-evolution baseline, confirming that the M23 sample is not significantly affected by evolutionary biases at current data quality. The analysis performed here leverages observational data, and in the future we can use the reconstructed light curve shown in \cite{Gupta2026JHEAp..5400684G}, in which we filled in gaps in the data to obtain more precise parameters of the 2D Dainotti relation. This work becomes even more relevant, as it shows the capability of the X-ray plateaus data to be used as cosmological standard candles. This work has been carried out leveraging 21 years of measurements taken consistently with the same satellite, the Swift mission \cite{Gehrels2004ApJ...611.1005G}, which is now de-orbited. GRBs reach $z \approx 5.6$ in the magnetar sample, extending the distance ladder well into the redshift gap between SNe~Ia and the CMB that is largely inaccessible to current probes. The combination of a physically motivated sample selection and a joint fitting methodology provides a clear path toward improved constraining efficiency as sample sizes grow. Forecasts suggest that with larger samples accumulated over the coming years, GRB plateaus may reach precision on $\Omega_m$ comparable to current SNe~Ia constraints~\cite{Dainotti:2022wli}. As discussed in \cite{Dainotti2027forecast}, within a few years GRBs might be able to constrain the dark energy equation of state $w$ parameter as precisely as Planck. The de-orbiting of the \textit{Swift} satellite makes it urgent to consolidate and extend the current dataset, particularly through optical follow-up campaigns that will provide complementary plateau measurements with improved photometric precision~\cite{Dainotti:2022jzv}. The methods developed here are directly applicable to such future samples and to the larger, higher-quality datasets expected from next-generation high-energy observatories.

\begin{acknowledgments}
K.F.D. acknowledges the support provided through the Research Excellence Programme administered by Xjenza Malta under project REP-2025-028, “Cosmological Artificial Intelligence” (CosmoAI). He also acknowledges the support and research facilities provided by the Malta College of Arts, Science and Technology (MCAST).
M.G.D. acknowledges the support of the JSPS Grant-in-Aid for Scientific Research (KAKENHI) (A), Grant Number JP25H00675. 
This article is also based upon work from COST Action CA21136 Addressing observational tensions in cosmology with systematics and fundamental physics (CosmoVerse) supported by COST (European Cooperation in Science and Technology). 
JLS would also like to acknowledge funding from ``Xjenza Malta'' as part of the ``Technology Development Programme'' DTP-2024-014 (CosmicLearning) Project. 
ALŁ acknowledges that the study was funded by the Research Support Module as part of the "Excellence Initiative - Research University" program at the Jagiellonian University in Kraków. 
N.F.  acknowledges  financial  support  from UNAM-DGAPA-PAPIIT  through  grant  IN112525.
The authors would like to acknowledge Purba Mukherjee for her contribution in early stages of this project. 
\end{acknowledgments}

\bibliography{references}

\end{document}